\documentclass[aps,article]{revtex4}%
\usepackage{amsfonts}
\usepackage{amsmath}
\usepackage{amssymb}
\usepackage{graphicx}%
\begin{document}
\preprint{ }
\title[Hierarchical game of life]{Complexity and hierarchical game of life}
\author{Ivan G\"{o}tz, Isaak Rubinstein, Eugene Tsvetkov and Boris Zaltzman\footnote{To
whom correspondence should be addressed}}
\affiliation{Blaustein Institute for Desert Research, Ben-Gurion University of the Negev,
Sede Boqer Campus 84990, Israel}

\begin{abstract}
Hierarchical structure is an essential part of complexity, important notion
relevant for a wide range of applications ranging from biological population
dynamics through robotics to social sciences. In this paper we propose a
simple cellular--automata tool for study of hierarchical population dynamics.

\end{abstract}
\maketitle

Hierarchical structure is an essential component of complexity, important
notion under intense current investigation with applications ranging from
biological population dynamics through robotics to group selection in social
sciences. In this paper we propose a simple cellular--automata tool for study
of hierarchical population dynamics.

There are numerous examples of complex organisms, such as slime moulds,
sponges and corals, which are in fact colonies of elementary organisms. These
colonies may in turn interact and form structures too. Thus, one may think of
a discrete hierarchy of structures in which those of each lower level form the
elementary components for next higher level. The natural question is: for a
given set of rules for element interaction at each level and that between the
levels, what is the effect of the hierarchy on the dynamics of the given
level; in particular, on the rate with which the evolution occurs at each
level, e.g. on the time of arrival at steady-state or quasi-steady-state
spacial structures, whenever such a state occurs and on the character of
structures which develop.

Answer to this question might be relevant for various aspects of population
dynamics in biological and social contexts, in particular, evolution genetics
and discrimination between altruistic and egoistic strategies, collective
robotics and software agent coordination, etc.

We propose an attempt to address this question by means of a simple cellular
automata model, similar to Conway's Game of Life \cite{1}, \cite{2} which we
call \textquotedblright Hierarchical Game of Life\textquotedblright\ \ (HGL).
Game of Life has been a popular tool for study of dissipative structures,
self-organized criticality and critical phenomena \cite{3}--\cite{10}. The
essence of HGL we propose is as follows:

The bottom level of the hierarchy is a periodic (toroidal) board with
$2^{n}\times2^{n}$ squares (cells), black or white, representing the basic
elements of this level. Four neighboring squares (\textquotedblright
children\textquotedblright) form a bigger (\textquotedblright
parent\textquotedblright) square, which represents the basic element of the
next level of the hierarchy, either black or white too. The total number of
levels, $n$, at most, is a specified parameter of the system. Elements of each
level play their color update game by the rules involving evaluation of other
elements in that level (the nearest neighbors, in the simplest case) and their
respective parent elements at a frequency which is a given fraction of that at
the lowest bottom level. During the \textquotedblright rest
intervals\textquotedblright, between the horizontal updates, elements of each
\textquotedblright parent\textquotedblright\ level (not the lowest bottom one)
are updated in accordance with their children elements state, and, possibly,
their current state. An example of simple \textquotedblright
near-diffusional\textquotedblright\ rules, almost identical for all levels,
except for the inevitable particularities of the lowest bottom and the highest
top level, is presented below, along with the results of the respective
simulations in a three-level hierarchy with a random initial distribution of
black and white cells. These results are compared with those for a 1-level
(non-hierarchical) system with the same \textquotedblright
diffusional\textquotedblright\ game rules.

We consider a three levels hierarchy ($1\leq n\leq3)$ of cells (see Fig.1).
Each cell, indexed by subscript $i$, is colored in black ($S_{i}^{n}=1$) or
white ($S_{i}^{n}=0$), where $S$ \ is a state function. Each cell has 8
neighbors $(k=1..8$, $k$ is the neighbor index of cell $i$) colored either in
black ($S_{i,k}^{n}=1$) or white ($S_{i,k}^{n}=0$). Cells of each level are
divided into the following two fixed populations of cells 'pragmatists' and
'romanticists'. The pragmatists cells are insensitive to hierarchy and play
their color update game by the following rules:

1. Horizontal pragmatists game (HPG): If there are more white neighbors than
black $\sum_{k=1..8}S_{i,k}^{n}<4,$ the cell $i$ assumes a white color
$S_{i}^{n}=0$, in the opposite case, $\sum_{i=1..8}S_{i,k}^{n}>4$ --- black
$S_{i}^{n}=1.$ If the number number of white and black neighbors is equal
$\sum_{i=1..8}S_{i,k}^{n}=4$ then the updated color is randomly selected with
an equal probability for black or white. The frequency of HPG is 1 on the
lowest bottom level, 2 on the intermediate level and 4 on the highest top
third level.

2. Vertical update (VU) occurs at each level at the intermediate time steps
between the horizontal updates at this level (odd time steps on the second
level, aliquant to four on the third level) with the same rules for
pragmatists and romanticists populations. The color of the each element of the
level is imposed by the dominant color of its four children ($S_{i}^{n,l},$
$l=1..4)$. If $\sum_{i=1..4}S_{i}^{n,l}<2$ then cell becomes white $S_{i}%
^{n}=0$, in the opposite case $\sum_{i=1..4}S_{i}^{n,l}>2$ -- black,
$S_{i}^{n}=1.$ If the number of white and black children is equal
$\sum_{i=1..8}S_{i}^{n,l}=2,$ then the color is randomly selected with an
equal probability for black or white.

The horizontal update game for the romanticists population occurs
simultaneously with HPG and is played by the following \textquotedblright top
impaired\textquotedblright\ diffusional rules:

3. Horizontal romanticists game (HRG): In this version, the horizontal update
of each cell is affected by its parent cell $\widetilde{S}_{i}^{n},$ which is
either black $\left(  \widetilde{S}_{i}^{n}=1\right)  $ or white $\left(
\widetilde{S}_{i}^{n}=0\right)  .$ If in the neighborhood of a cell the black
color is not dominant $\sum_{i=1..8}S_{i}^{n}\leq4$ and the parent's color is
white $\widetilde{S}_{i}^{n}=0,$ then the cell becomes white. If the
neighborhood is dominated by the black color $\left(  \sum_{i=1..8}S_{i}%
^{n}>4\right)  $ whereas the parent cell's color is white $\widetilde{S}%
_{i}^{n}=0$ $,$ the updated color is randomly selected with an equal
probability for the black or white. If in the neighborhood of a cell the white
color is not dominant $\left(  \sum_{i=1..8}S_{i}^{n}\geq4\right)  $ and the
parent's color is black $\left(  \widetilde{S}_{i}^{n}=1\right)  ,$ the cell
becomes black$.$ Finally, if the neighborhood is white dominated $\left(
\sum_{i=1..8}S_{i}^{n}<4\right)  $ and the parent's color is black $\left(
\widetilde{S}_{i}^{n}=1\right)  ,$ the updated color of the cell is randomly
selected with an equal probability for the black and white. The frequency of
HRG is 1 on the lowest bottom level, 2 on the intermediate level and 4 on the
highest top third level.

Initial distribution of the white and black cells in the bottom level, and
distribution of pragmatists and romanticists at all levels are random with
probability $P_{1}$ for a cell to have initially a black color and probability
$P_{2}$ to belong to the romanticists kind. Once assigned, romanticism or
pragmatism are permanent characteristics of a given cell. Naturally, the cells
of the top level (the third level, in our case) have no parents , thus all of
this top level are assumed pragmatists.

To simulate this hierarchy we used the attached $C\!\mathit{++}$ code (see
website \textsl{www.math.bgu.ac.il%
%TCIMACRO{\TEXTsymbol{\backslash}}%
%BeginExpansion
$\backslash$%
%EndExpansion
\symbol{126}borisz/GameLife/Simlife.exe}). The main control parameter in our
simulations for this example was the romanticists ratio in the population
$\eta=\frac{\text{number of romanticists cells}}{\text{total number of cells
in the population.}}.$ For simplicity we set $\eta$ the same for all levels. A
series simulation were run for different values of $\eta,$ $0\leq\eta\leq1$
(See Fig. 2a--d)$.$ We began with a purely pragmatic population $\eta=0$. In
this case the population dynamics was purely diffusional, and the system
converged in a rather long but finite time to one of the following two
steady-state or quasi-steady-state configurations:

1. The entire population either black or white;

2. Field of a uniform color, split into two parts by a straight strip of a
different color. In the latter case the strip boundary is either a stationary
straight line or a \textquotedblright running zipper\textquotedblright: wave
of cells changing color in time\ (Fig. 2a).

Upon the increase of $\eta$ these configurations transform into a
quasi-steady-state labyrinth patterns with possible islands (see Fig. 2b--d).
The inner cells of these patterns preserved their color in time, whereas at
the boundary the cell color may keep changing. We define the complexity
$\alpha\geq0$ of the labyrinth patterns and the degree of unsteadiness of the
system $\beta\geq0$ as the following ratios:%
\[
\alpha=\frac{\text{number of cells on the black-white boundary }}{\text{total
number of cells}}%
\]%
\[
\beta=\frac{\text{number of boundary cells changing color }}{\text{total
number of cells}}.
\]
For long times, the deviations of $\alpha$ and $\beta$ from the respective
average values become small.

The increase of the romanticism ratio $\eta$ results in the increase of the
population's complexity and unsteadiness, accompanied by the increase of the
number of black and white islands (Fig. 2c,d, Fig. 3). Finally, we define the
relaxation time $T>0$ as the number of elementary time steps upon which the
ratio of black cells to the total number of cells becomes constant with a
given accuracy (1\%, in our case). In Fig. 4 we present the dependence of the
relaxation time $T$ on the romanticism ratio $\eta$. The main observations in
this example are as follows:

The increase of romanticism ratio in the population and the related
sensitivity to the hierarchy results in:

1. Formation of the labyrinth and islands pattern and the related increase of
system's complexity;

2. Increase of the population unsteadiness;

3. Sharp decrease of the relaxation time. This observation of accelerated
convergence to a quasi-steady-state within a hierarchy might be of potential
importance for modeling the evolution of systems from simple to complex forms.

We wish to emphasize that the simulation above is just an example of the
hierarchy effect on the population dynamics. The rules of the vertical and
horizontal games (as natural as they may seem) are rather arbitrary and so are
their effects. Our objective was only to present a simple tool for modeling
the relation between the hierarchy in the system and its complexity.

The following remark is due: a hierarchy arises in any statistical system with
cluster formation and phase separation. How reasonable is the postulation \ of
element interaction rules at each hierarchy level? Rather, should they not be
worked out from the elementary interactions at the lowest bottom level of the
hierarchy? The answer to this question seems to be negative. Indeed for a
sufficiently high hierarchical pyramid with an arbitrary discretization into a
few separate levels information is lost upon the transition from the lower to
higher levels. Thus, inferring the \ interaction rules for the elements of the
top level from those of the bottom one appears as reasonable as inferring
individual or national psychology from the properties of elementary particles.
Rather, a reasonable identification of the interaction rules for each
hierarchical level as the basis for investigation of the effect of hierarchy
upon the dynamics of population seems to be a more practical approach.

\newpage

\subsection{{\protect\LARGE Legend to Figures}}

\begin{description}
\item Fig.1. \ Sketch of a three level hierarchy

\item[ ] Fig.2 \ Quasi-steady-state configurations for the bottom level of a
three levels hierarchy for different fraction of 'romanticists' cells
($\eta=0,$ $0.05,$ $0.2,$ $1)$:

\item a) Quasi-steady-state strip: left boundary -- steady state, right
boundary -- \textquotedblright\ running zipper\textquotedblright. Purely
diffusional rules, no sensitivity for hierarchy for the bottom level
($\eta=0).$

\item b) 5\% 'romanticists' cells fraction on all levels: 'romanticists' are
darker on the black background and grey on the white background.

\item c) 20\% 'romanticists' cells fraction on all levels: 'romanticists' are
darker on the black background and grey on the white background.

\item d) 100\% 'romanticists' populations on all level ($\eta=1).$

\item Fig.3. Complexity $\alpha$ and degree of unsteadiness $\beta$ as
functions of the 'romanticists' fraction $\eta.$

\item Fig.4. Dependence of time of relaxation to the quasi-steady-state ($T$)
on the 'romanticists' fraction $\eta.$
\end{description}

\end{document}